\documentclass[12pt]{article}
\usepackage{psfig,float}
\begin{document}

\begin{center}
{\Large{\bf Delta excitation in $K^+$-nucleus collisions}}
\end{center}

\vspace{1cm}

\begin{center}
{\large{ J. A. Oller$^1$,
E. Oset$^1$, A. De Pace$^2$ and  P. Fern\'andez de C\'ordoba$^3$}}
\end{center}

\vspace{0.4cm}

{\it $^1$ Departament de F\'{\i}sica Te\`orica, and IFIC
Centro Mixto Universidad de Valencia-CSIC;
46100 Burjassot (Valencia), Spain.

 $^2$ Istituto Nazionale di Fisica Nucleare, Sezione di Torino,
via P. Giuria 1, I-10125 Torino, Italy.

$^3$ Departamento de Matem\'atica Aplicada, Universidad Polit\'ecnica de
Valencia, Spain.
}

\vspace{1cm}

\begin{abstract}
{\small{We present calculations for $\Delta$ excitation in the
($K^+,K^+)$ reaction in nuclei. The background from quasielastic
$K^+$ scattering in the $\Delta$ region is also evaluated and shown
to be quite small in some kinematical regions, so as to
allow for a clean identification of the $\Delta$ excitation strength.
Nuclear effects tied to the $\Delta$ renormalization in the nucleus 
are considered and the reaction is shown to provide new elements to 
enrich our knowledge of the $\Delta$ properties in a nuclear medium.}}
\end{abstract}

\section{Introduction.}

Delta excitation in nuclei has been a topic of permanent interest
and it has been studied in connection with pion nucleus collisions 
\cite{1}, photonuclear reactions \cite{2,3,4,5}, electron scattering on nuclei
\cite{5a,5b,6,7} nuclear reactions induced by protons or light nuclei
\cite{7a,8,9,10}, neutrino induced reactions \cite{11,12}, etc.

In all these reactions the $\Delta$ excitation proceeds in a different
way: sometimes it is excited by a spin-isospin longitudinal source
(pions), other times by a transverse source (photons), and in other cases
by a mixture of both. Also, the range of energy and momentum used to 
excite the $\Delta$ varies from one case to another. Differences also
appear in the regime of nuclear densities explored. In some reactions
the $\Delta$ is more neatly excited than in others where background
terms are important and, often, distortions of the strongly
interacting particles
involved in the reaction lead to $\Delta$ shapes that differ appreciably
from each other. All these differences, however, serve to enrich our 
knowledge of the  $\Delta$ properties in a nuclear medium and of its
coupling to the nuclear components.

Delta excitation in $K^+$ nuclear reactions has not yet been explored
and clearly deserves some attention in view of its complementarity with
respect to other reactions mentioned above. 

The $K^+$ is a meson belonging, like the pion, to the octet of pseudoscalar
mesons. However it has peculiar features. In a sense, the small $K^+ N$ 
cross sections  allow the kaons to explore inner regions of the 
nucleus, while pion nuclear reactions are usually more peripheral. 
Another big difference is the fact that the pion can be absorbed by one
nucleon to give the $\Delta$, while this is not possible with the
$K^+$ due to its strangeness. One can also not excite strange
baryons (of negative strangeness) with the $K^+$. Hence the $K^+$ in this 
case can only release some momentum and energy and keep traveling
as a $K^+ $ (or $K^0$). In this sense the 
$\Delta$ excitation induced by $K^+$ is similar to the proton induced
one in $(p,p')$ or $(p,n)$ reactions, with the difference that
in the $K^+$ case the $\Delta$ is excited only with a transverse source
as we shall see.

The modifications of the $\Delta$ properties in a nuclear medium have been
the object of much theoretical attention \cite{13,14,15,16,17,18}.
Also early empirical studies of pion-nucleus scattering lead to
parametrizations of the $\Delta$ spreading potential \cite{19},
although caution has to be exerted to compare to theoretical 
models because the empirical spreading potential incorporates elements which
in some theoretical models are part of the $\Delta h$ interaction
\cite{1}. 

Experiments on $K^+$-nucleus scattering in the $\Delta$
excitation region would bring additional information by means of which
to test our present understanding of the $\Delta$ properties in nuclei
and enrich it.

Experiment on inclusive $K^+$-nucleus scattering are already available
\cite{20,21}, but they restrict themselves to the quasielastic
excitation region. These data have proved useful in order to learn about
the strength of the residual nuclear forces \cite{22}, since they have offered
new information with respect to the one obtained from electron scattering 
at low momentum transfers \cite{23}. The extension
of this work to the $\Delta$ excitation region, passing through the dip
region should be most useful. We should recall that the dip region
has been a permanent theoretical problem in inclusive electron scattering
and only the recent thorough many body calculation of ref. \cite{7} has been
able to provide a fair description so far. Given the different 
dynamics in $K^+$-nucleus scattering with respect to electrons,
we anticipate that this region should pose a challenge to theory.

The elementary $K^+ N \rightarrow K \Delta$ reaction has not been
much studied but there are data for $K^+ p \rightarrow K N \pi$ in 
several charge channels which clearly indicate the contribution from
$\Delta$ excitation \cite{24,25}. A recent study of this reaction using
the terms from chiral Lagrangians plus $\Delta$ excitation, has
been performed in \cite{26} and this provides us with the elementary
information needed to tackle the nuclear problem. The other important
ingredient is the $\Delta$ selfenergy
in the nuclear medium, which we take from ref. \cite{15}. This selfenergy
has been tested in elastic pion nucleus scattering \cite{27} and in
quasielastic, single charge exchange, double charge exchange and 
absorption of pions in nuclei  \cite{28}. It has also been
tested in photonuclear reactions \cite{5} and electronuclear reactions
\cite{7}, and in all cases a good description of the data around the
resonance region was found. With these ingredients at hand we tackle
now the $K^+$ nucleus inclusive scattering around the $\Delta$ 
region.

\section{The model.}

Following the developments in photonuclear and electronuclear 
reactions \cite{5,7} we evaluate the selfenergy of a $K^+$ in
nuclear matter and from there the cross section in nuclei via the local 
density approximation.

\begin{figure}[h]
\centerline{
\protect
\hbox{
\psfig{file=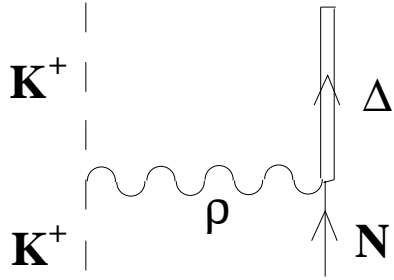,width=.42\textwidth,angle=-90}}}
\caption{$\Delta$ excitation term mediated by $\rho$-exchange in the
$K^+N\rightarrow K^+\Delta$ reaction.}
\end{figure}

The elementary model of \cite{26} for $K^+ N \rightarrow K^+ \Delta$ is
depicted in fig. 1. The model consists of $\rho$ exchange between the
kaon and the baryonic components. The two necessary ingredients are the
$K^+ K^+ \rho$ coupling and the $\rho N \Delta$ coupling, which we
take from \cite{26} where a fit to the data was performed. We have for
$\rho^0 \rightarrow K^+ K^- $
\begin{equation}
- i \delta H_{\rho K^+ K^-} = - i \tilde{f}_\rho \epsilon^\mu_\rho 
[p_{K^+} -
p_{K^-}]_\mu
\end{equation}
and for $\rho^0 N \rightarrow \Delta$ the vertex function ($\vec{q}
\equiv \rho $ momentum)
\begin{equation}
- i \delta \tilde{H}_{\rho^0 N \Delta} = 
\sqrt{\frac{2}{3}} \frac{f^*}{m_\pi} \sqrt{C_\rho} (\vec{S}\,^{\dagger}
 \times
\vec{q}) \cdot \vec{\epsilon}_\rho ,
\end{equation}
where $\epsilon^\mu_\rho$ is the polarization vector of the $\rho$ and 
$S^\dagger$ the
spin transition  operator from spin 1/2 to 3/2. The coefficient $\sqrt{2/3}$
is an isospin coefficient. In addition we use a monopole form
factor for the $\rho N \Delta$ vertex of the type
\begin{equation}
F_\rho (q) = \frac{\Lambda^2 - m_\rho^2}{\Lambda^2 - q^2} ,
\end{equation}
with $\Lambda = 2$ GeV. By fixing $C_\rho = 2$ and using the
standard value $f^{*2}/4 \pi = 0.36$, the fit to the data in
\cite{26} gave a value $\tilde{f}_\rho = 4.2$, 30$\%$ higher than the
expected $SU (3)$ value $\tilde{f}_\rho = f_\rho/2 = 3.1 $ \cite{29,30}.
This value, however, is imposed by our choice of the $\rho N \Delta$ 
coupling, where we rely again on $SU (6)$ symmetry to relate it to the
empirical $\rho NN$ coupling used in \cite{5}. 

\begin{figure}[h]
\centerline{
\protect
\hbox{
\psfig{file=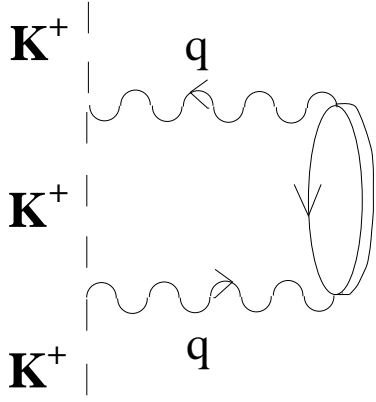,width=0.35\textwidth,angle=-90}}}
\caption{Selfenergy of $K^+$ associated with an intermediate $\Delta h$
excitation.}
\end{figure}

Next step is to evaluate the $K^+$ selfenergy in nuclear matter where the
intermediate state is $K^+$ and a $\Delta h$ excitation. This selfenergy
diagram is depicted in fig. 2. By using the sum over $\Delta$
spins,
\begin{equation}
\sum_{M_s} S_i |M_s > < M_s | S^\dagger_j = \frac{2}{3} \delta_{ij} 
- \frac{i}{3}
\epsilon_{ijk} \sigma_k ,
\end{equation}
and taking into account that the three momenta of the $\rho N \Delta$
coupling must be taken in the $\Delta$ CM frame, we can write in terms
of the $K^+ A$ Lab frame momenta the kaon selfenergy as
\begin{eqnarray}
\Pi (k) & =&  i \int \frac{d^4 q}{(2 \pi)^4}
D^2_\rho (q) (\frac{f^*}{m_\pi })^2
C_\rho \tilde{f}\,^2_\rho \tilde{U}_\Delta (q)\\ \nonumber
&& \times \frac{16}{9} (\frac{M}{M_I})^2 (\vec{k} \times \vec{k}\,')^2
D_{K^+} (k - q) F^2_\rho (q) ,
\end{eqnarray}
where $M$ is the nucleon mass, $M_I$ the invariant mass of the
$\Delta$, $M_I^2 = p^{02}_\Delta - \vec{p}_\Delta^2$, 
$D_{K^+}$ and $D_\rho$ are the $K^+$ and $\rho$ propagators respectively and 
$\tilde{U}_\Delta (q)$ is the $\Delta h$ Lindhard function with the
normalization
\begin{equation}
\tilde{U}_\Delta (q) = \rho \frac{1}{\sqrt{s} - M_\Delta +
i \Gamma(s)/2} ,
\end{equation}
with $\rho$ the nuclear density.

The step from $\Pi (k)$ to a nuclear cross section is readily done by
recalling that the reaction probability per unit time is $(2 \omega
V_{opt} \equiv \Pi)$ 
\begin{equation}
\Gamma = - 2 Im V_{opt} = - \frac{1}{\omega} Im \Pi (k) ,
\end{equation}
with $\omega$ the kaon energy. The probability of reaction per unit
length is then $- Im \Pi/k$ and hence the contribution of an
element of volume to the cross section is
\begin{equation}
d \sigma = - \frac{1}{k} Im \Pi (k) d^3 r .
\end{equation}

The local density approximation comes now into action since $\Pi (k)$ is a 
function of $\rho$, the nuclear density, and then the cross section
in a finite nucleus becomes
\begin{equation}
\sigma = - \frac{1}{k} \int d^3 r Im \Pi (k, \rho (\vec{r})) .
\end{equation}

One must now evaluate $Im \Pi$ from eq. (5), which is readily done using
Cutkosky rules, placing on shell the intermediate  states of the
selfenergy diagram. Technically on has
\begin{eqnarray}
\Pi (k) &\rightarrow & 2 i Im \Pi \\ \nonumber
\tilde{U}_\Delta (q) &\rightarrow & 2 i \theta (q^0) Im \tilde{U}_\Delta (q) \\
\nonumber
D_{K^+} (k - q) &\rightarrow &2 i \theta (k^0 - q^0) Im D_{K^+} (k - q) =\\
&& 2 i  \frac{1}{2 \omega (\vec{k} - \vec{q})}
(- \pi) \delta (k^0 - q^0 - \omega (\vec{k} - \vec{q})) .
\end{eqnarray}
This allows us to write the $K^+$ differential cross section as
\begin{eqnarray}
\frac{d \sigma}{d \Omega ' d \omega '}& = & \int  \frac{d^3 r}{(2 \pi)^3} 
\frac{k'}{k} \frac{8}{9} (\frac{f^*}{m_\pi})^2 
C_\rho \tilde{f}_\rho^2 (-) Im \tilde{U}_\Delta (k - k')\\ \nonumber
&&\quad\times (\frac{M}{M_I})^2 (\vec{k} \times \vec{k}')^2 D_\rho (q)^2
F_\rho^2 (k - k) ,
\end{eqnarray}
with $D_\rho (q) = (q^2 - m_\rho^2)^{-1}$.

So far we have not introduced $\Delta$ selfenergies into the scheme. 
There is also another physical effect that must be taken into account
which is the distortion of the $K^+$ waves.

The $\Delta$ selfenergy is readily introduced adding $\Sigma_\Delta$ from
ref. \cite{15} to the $\Delta$ mass in $\tilde{U}_\Delta (q)$, including
Pauli corrections to the $\Delta$ width.

\begin{figure}[h]
\centerline{
\protect
\hbox{
\psfig{file=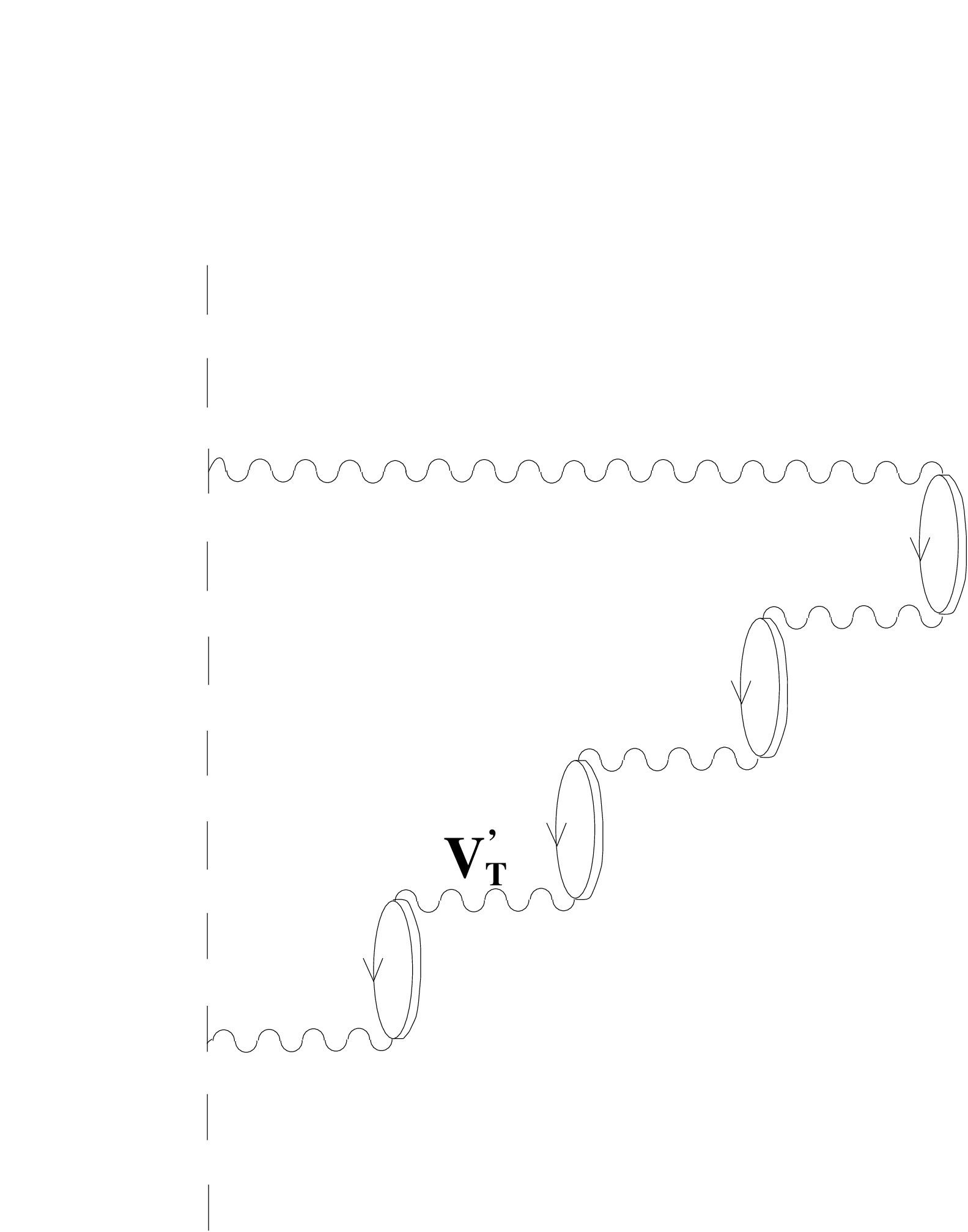,width=0.52\textwidth,angle=-90}}}
\caption{Contribution to the $K^+$ selfenergy from Tamm-Dancoff propagation 
of $\Delta h$ states.}
\end{figure}

At the same time one can introduce  corrections from the RPA propagation
of $\Delta h$ in the  medium to account for the diagrams of the
type depicted in fig. 3, where backward going $\Delta h$ excitations are
omitted since they are negligible in the $\Delta$ region. This is also
accomplished technically in a very easy way 
\cite{5} by substituting $\Sigma_\Delta$ by 
\begin{equation}
\Sigma_\Delta \rightarrow \Sigma'_\Delta = \Sigma_\Delta +
\frac{4}{9} (\frac{f^*}{m_\pi})^2 V'_T \rho ,
\end{equation}
where $V'_T$ is the transverse part of the spin-isospin interaction,
\begin{equation}
V'_T = \frac{\vec{q}\,^2}{q^2 - m_\rho^2} C_\rho F_\rho^2 (q) + g' ,
\end{equation}
and $g'$, the Landau-Migdal parameter, is taken as $g' = 0.6$.

The next correction is the distortion of the kaons. This requires some
thought because the $K^+$ is distorted only by quasielastic
collisions or conversion into $K^0$. In the latter case the $K^+$ 
disappears after one collision (although it can be generated again in 
a second collision), but in the quasielastic collisions the $K^+$ remains,
although changing direction and energy. The conventional use of a 
$K^+$-nucleus optical potential removes from the $K^+$ flux all events where
there is a quasielastic collision or $K^0$ conversion. However, for small
angles of the emerging $K^+$ this procedure is numerically accurate
since the contribution of two step processes, one quasielastic and the other
one the $N \Delta$ tansition, is negligible at small angles. This has
been found as a general rule in hadronic collisions \cite{31}, in the 
$\Delta$ excitation with the $(^3 He, t)$ reaction \cite{32,33} and in
$K^+$ quasielastic scattering \cite{22}, much closer to the problem
we are dealing with.

Since we are going to deal with small $K^+$ angles, we shall then
use distorted waves for the $K^+$ and the same assumption of small angles
allows us to use the eikonal approximation. In this case we must 
multiply the cross section of eq. (11) by the distortion factor
$D (k, k', \vec{r})$ given by
\begin{equation}
D (k, k', \vec{r}) = exp \left( \int_{- \infty}^z \sigma_{KN}^{(1)} \rho (
\vec{b}, z') dz' + \int_z^\infty \sigma_{KN}^{(2)} \rho (\vec{b},
z') dz' \right) ,
\end{equation}
where $\vec{b}$ is the impact parameter corresponding to the point
$\vec{r}$ and $\sigma_{KN}^{(1)}, \sigma_{KN}^{(2)}$ are the
$K^+ N$ cross sections of the incoming and outgoing $K^+$
respectively, which we take from \cite{34}.

Summarizing, our final formula for the cross section is given by
eq. (11) multiplying the expression by the distortion  factor of eq. (14)
and substituting $M_\Delta$ by $M_\Delta + \Sigma'_\Delta$ in
$\tilde{U}_\Delta (q)$ of eq. (6), with $\Sigma'_\Delta$ given by
eq. (12).

We present results in the next section.

\section{Results and discussion.}

\begin{figure}[h]
\centerline{
\protect
\hbox{
\psfig{file=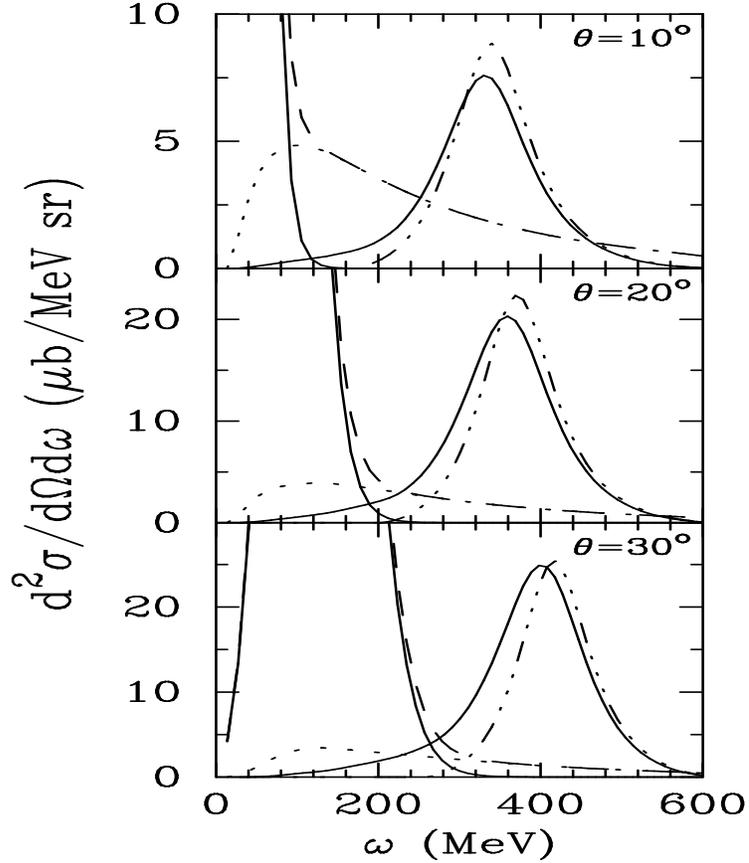,width=0.7\textwidth,height=0.5\textheight}}}
\caption{Double differential cross sections for $K^+$ scattering on 
$^{12}$C at 1 GeV/c and for three different angles. In the $\Delta$ region
results for the free $\Delta$ (dot-dot-dot-dashed) and for the medium-modified
$\Delta$ (solid) are displayed. The dashed line represents the total 
quasielastic background due to one-step (solid) and two-step (dotted) 
collisions.}
\end{figure}

In fig. 4 we show differential cross sections for $\Delta$ production
for $k = 1$ GeV/c and three different angles for $^{12}$C. At the
same time we calculate the background from quasielastic $K^+$ 
collisions in the same region, coming from one and two steps, as
discussed in \cite{22}. Since our aim is to single out kinematical regions 
where this background can be expected to be negligible, these latter 
calculations have been performed with some simplifications, that is the use
of harmonic oscillator states (instead of Woods-Saxon), the
omission of the RPA corrections (which were found relevant only on the 
left hand side of the quasielastic peak) and the  neglect of the 
width of the $ph$ states. We can see that at $\theta = 10^0$ there is a 
substantial background below the $\Delta$ peak coming from two-step 
quasielastic collisions. The figure also shows the effect of the
$\Delta$ selfenergy and the $\Delta h$ interaction in the transverse
channel (addition of $\Sigma'_\Delta$ to $M_\Delta$). There is a small
shift of the peak to smaller excitation energies, a moderate decrease of the 
strength at the $\Delta$ peak and some increased strength at lower
excitation energies, which comes as a consequence of the $\Delta$
coupling to $ph$ components, i. e., the decay mode of the $\Delta$ in the
nucleus, $\Delta N \rightarrow NN$. We can see this strength more visible
 at bigger angles $\theta = 20^0, 30^0$. For these latter angles the 
quasielastic background is relatively smaller, which makes it easier to
identify the $\Delta$ excitation strength.

\begin{figure}[h]
\centerline{
\protect
\hbox{
\psfig{file=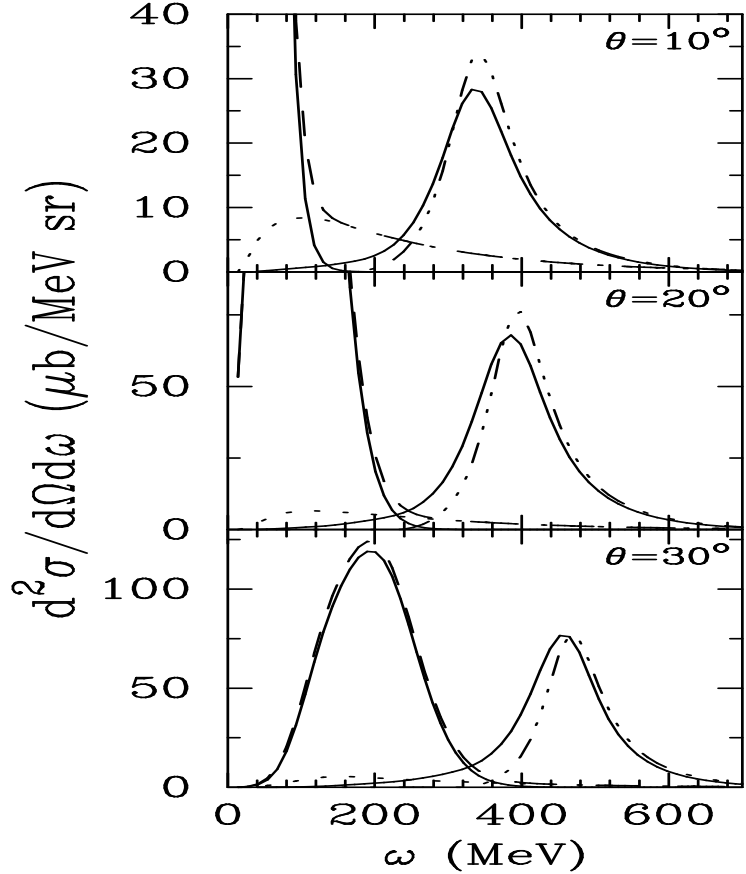,width=0.7\textwidth,height=0.5\textheight}}}
\caption{As in Fig.4, but for $K^+$ at 1.25 GeV/c.}
\end{figure}

In fig. 5 we show the same results for $k = 1.25 \; GeV$. The
qualitative features here are similar to those in fig. 4, only the 
relative strength of the $\Delta$ excitation with respect to the 
quasielastic one is bigger.

\begin{figure}[h]
\centerline{
\protect
\hbox{
\psfig{file=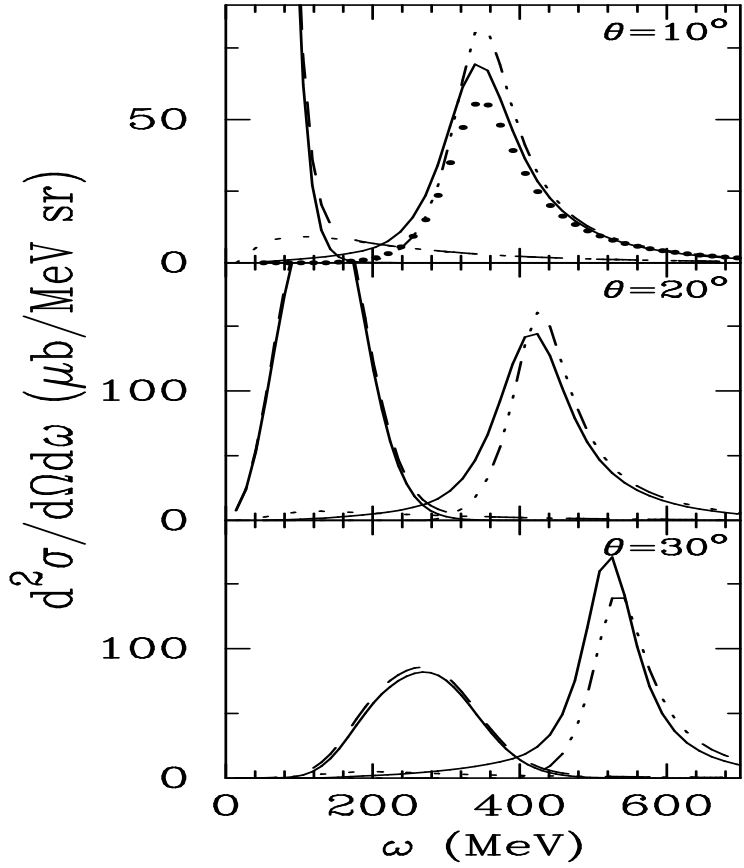,width=0.7\textwidth,height=0.5\textheight}}}
\caption{ As in Fig.4, but for $K^+$ at 1.5 GeV/c.
In the top panel, the amount of $\Delta$ strength in the medium due to pionic
decay is also shown (heavy dots).}
\end{figure}

In fig. 6 we show the results for $k = 1.5 \; GeV$. Once again the
features are similar to those in the former figures and 
the strength of the $\Delta$ excitation with respect to the
quasielastic one is even higher.

At the angle $\theta = 30^o$ the $\Delta $ strength is bigger than the 
quasielastic one, but the quasielastic contribution has a wide bump
that induces an appreciable background below the $\Delta$ peak.

The effects of the $\Delta$ selfenergy in the medium might look
moderate by comparing the solid and dash-dotted lines in figs. 4-6.
However, the medium effects are far more relevant than these
two lines might indicate. Indeed, in the case of a free $\Delta$,
the width is fully associated to the pionic decay of the $\Delta$
 while in the nuclear medium the width is associated to pion
emission and $ph$ excitations and only part of the $\Delta$ strenght
of the figure goes into pion emission. This can be made quantitative
by recalling  the form for the $\Delta$ selfenergy from \cite{15}. We
have
\begin{eqnarray}
&&\tilde{U}_{R, \Delta} (q) = 
 \rho \\
&& \times \frac{1}{\sqrt{s} - M_\Delta + i \tilde{\Gamma}/2 - Re  
\Sigma_\Delta
- \frac{4}{9} (\frac{f^*}{m_\pi})^2 V'_T \rho + i 
C_Q (\frac{\rho}{\rho_0})^\alpha + i C_{A2} (\frac{\rho}{\rho_0})^\beta
+ i C_{A3} (\frac{\rho}{\rho_0})^\gamma } , \nonumber
\end{eqnarray}
where $\tilde{U}_{R,\Delta}$ is the $\Delta h$ Lindhard function
incorporating  the selfenergy corrections. In eq. (15) $\rho_0$ is the normal
nuclear matter density, $\tilde{\Gamma}$ is the Pauli blocked width and
$C_Q, C_{A2}, C_{A3}$ are coefficients parametrized
in \cite{15} such that their corresponding terms are associated to $\Delta$
pionic decay  $(C_Q)$, $2 p \,1 h$ decay  $(C_{A2})$ and
$3 p \, 2h  $ decay $(C_{A3})$.

The strength of the $\Delta$ decaying into pions is associated to $
\tilde{\Gamma}$ and the $C_Q$ term and we can write
\begin{eqnarray}
&& Im \tilde{U}_{R,\Delta} (q) = - \rho \\
&& \times \frac{
\tilde{\Gamma}/2 + C_Q (\frac{\rho}{\rho_0})^\alpha
+ C_{A2} (\frac{\rho}{\rho_0})^\beta + C_{A3} (\frac{\rho}{\rho_0})^\gamma }
{(\sqrt{s} - M_\Delta - Re \Sigma_\Delta - \frac{4}{9} (\frac{f^*}{
m_\pi})^2 V'_T \rho)^2 + (\frac{\tilde{\Gamma}}{2} + C_Q 
(\frac{\rho}{\rho_0})^\alpha
+ C_{A2} (\frac{\rho}{\rho_0})^\beta + C_{A3} (\frac{\rho}{\rho_0})^\gamma
)^2 } . \nonumber
\end{eqnarray}
With this separation and bearing in mind the meaning of 
Cutkosky rules, if we take the first two terms in the numerator of
eq. (16), the resulting strength will go into primary pion emission,
while the one coming from the last two terms will go into nucleon
emission.

We have thus isolated the pionic decay content of the $\Delta $ strength
and show it in fig. 6 at $\theta = 10^0$.
This strength is only about 70$\%$ of the
corresponding one for a free $\Delta$ and the reduction is not due to the
Pauli blocked width but to the competition of the other
$\Delta$ decay channels. Indeed, in the absence of
$ph$ $\Delta$ decay channels, $Im \tilde{U} \sim \tilde{\Gamma}^{-1}$,
and with a reduced $\tilde{\Gamma}$ width, the $\Delta$ peak
would increase rather that the opposite, while at the same time the
resonance shape would become narrower.

We should also point out that this pionic content refers to the first
step of the reaction, before there is any final state interaction.
Recall that in our local density formula we are producing the pions
in an element of volume $d^3 r$. In their way out, part of these
pions will be reabsorbed and will show up as particle emission. In a 
nucleus like $^{12}C$, about 30$\%$ of these pions are reabsorbed
\cite{12,35}, so that finally only about 1/2 of the original strength 
assuming a free $\Delta$ goes into pion emission.

It would be interesting to perform some coincidence measurements 
where pions would be detected together with the $K^+$.

We should also recall that the present reaction has other added
advantages over the $(^3 He,t)$ reaction which has been
thouroughly studied. Indeed, the $\Delta$ information on that 
reaction is essentially limited to $0^0$, since the cross section falls
by about two orders of magnitude when going to about $5^0$ 
and the shape of the $\Delta$ resonance is essentially lost \cite{36}.
Here, on the contrary, the cross section remains sizeable up to angles
of about $30^0$ and more. This offers a wider spectrum of
excitation energies and momenta by means of which to study the
$\Delta$ excitation.

\section{Conclusions.}

We have evaluated the cross section for inclusive $(K^+, K^+)$
scattering in nuclei around the $\Delta$ resonance region. These are
the first evaluations for a reaction on which there are no data yet,
but they could be obtained as a continuation of the recent
experimental program in the quasielastic region \cite{20,21}.

The cross sections obtained are sizable, and the mixture
with the quasielastic tail is sufficiently small in some regions
to allow for a clean separation of the $\Delta$ excitation and the
nuclear effects associated to it. The present study should stimulate
such measurements that surely will contribute to enrich
our knowledge of resonance renormalization in nuclei, which is
a subject of continuous debate.

\vspace{3cm}

{\bf Acknowledgments.} We would like to acknowledge support from 
the EU network CHRX-CT93-0323, and the hospitality of the
Universities of Valencia and Torino. One of us, J.A.O. wishes to
acknowledge support from the Generalitat Valenciana. This work is
partly supported by DGICYT contract number PB96-0753.

\end{document}